\documentclass[pra,twocolumn,superscriptaddress,notitlepage]{revtex4-1}

\usepackage{graphicx,amsmath,amsfonts,amssymb,upgreek,txfonts,color}
\usepackage[colorlinks,linkcolor=blue,citecolor=blue,urlcolor=blue,breaklinks=true]{hyperref}

\begin{document}

\title{Effects of pressure on suspended micromechanical membrane arrays}

\author{Andreas Naesby}
\affiliation{Department of Physics and Astronomy, University of Aarhus, DK-8000 Aarhus C, Denmark}
\author{Sepideh Naserbakht}
\affiliation{Department of Physics and Astronomy, University of Aarhus, DK-8000 Aarhus C, Denmark}
\author{Aur\'{e}lien Dantan}
\email[Corresponding author: ]{dantan@phys.au.dk}
\affiliation{Department of Physics and Astronomy, University of Aarhus, DK-8000 Aarhus C, Denmark}

\date{\today}

\begin{abstract}
The effects of pressure on micromechanical air-filled cavities made by a pair of suspended, parallel silicon nitride membranes are investigated in the free molecular and quasi-molecular regimes. Variations of the fundamental drummode mechanical resonant frequencies and damping with air pressure are determined by means of optical interferometry. A kinetic damping linear friction force and a positive resonant frequency shift due to the compression of the fluid between the membranes are observed to be proportional with pressure in the range 0.01-10~mbars. For resonators with near-degenerate modes hybridization of the modes due to this squeeze film effect is also observed and well accounted for by a simple spring-coupled oscillator model.
\end{abstract}

\maketitle

Due to their high frequency, low mass and the possibilities for electric, magnetic or optical activation, suspended micromechanical resonators are ubiquitous in force sensing, mass detection, displacement measurements, atomic force microscopy or signal processing~\cite{Bao2000,Ekinci2005,Aspelmeyer2014}. At low pressures the damping of vibrating nano- and microstructures is essentially determined by material properties and boundaries~\cite{Joshi2014}, allowing, e.g., for high sensitivity mass measurements~\cite{Ilic2004,Yang2006}. As the pressure is increased fluid-structure interactions typically become the dominant source of damping and limit the quality factors of the mechanical resonances. For nano- and microscale devices structures surrounding the resonator which constrain the gas flow and, for small gaps, can lead to \textit{squeeze film} effects which appreciably modify mechanical resonance frequencies and damping~\cite{Bao2007}. Squeeze film effects have been investigated for a variety of resonators~\cite{Andrews1993a,Darling1998,Bao2007,Verbridge2008,Okada2008,Suijlen2009,Stifter2012} and can be exploited, {\it  e. g.}, in pressure sensing~\cite{Andrews1993b,Southworth2009,Kainz2014,Kumar2015} or ultrasonic transduction~\cite{Hansen1999,Wygant2008,KhuriYakub2011} applications.

In this letter we experimentally investigate the effect of pressure on the mechanical properties of a pair of suspended, parallel silicon nitride membranes forming few micron-long cavities. The variations of the fundamental square drummode frequencies and quality factors are determined by optical interferometry for pressures between $10^{-5}$ and 10 mbars, and compared with those of  single membrane resonators. For resonators with well-separated mechanical resonance frequencies damping is found to be essentially of kinetic origin when the pressure increases, whereas the compressibility of the fluid is observed to add a positive mechanical spring constant which is proportional with pressure. In the case of membranes with near-degenerate mode frequencies hybridization of the vibrational modes of both membranes is observed. The experimental observations are in excellent agreement with theoretical predictions taking into account squeeze film effects in the rarefied air regime.

 Since such suspended membranes have demonstrated record-high mechanical quality factors~\cite{Tsaturyan2017}, very low optical losses, excellent optomechanical properties~\cite{Thompson2008} and capabilities for low-noise electro-opto-mechanical signal transduction and detection~\cite{Bagci2014,Andrews2014,Fink2016}, gas-coupled multiple membrane resonators, such as the ones investigated in this work, are promising for realizing integrated micromechanical cells for pressure sensing and ultrasound transduction applications~\cite{Leinders2015}, spectroscopy~\cite{Jensen2016} or for collective and hybrid optomechanics~\cite{Xuereb2012,Xuereb2014,Xuereb2015,Dantan2014,Moeller2017}.

\begin{figure}
\includegraphics[width=0.8\columnwidth]{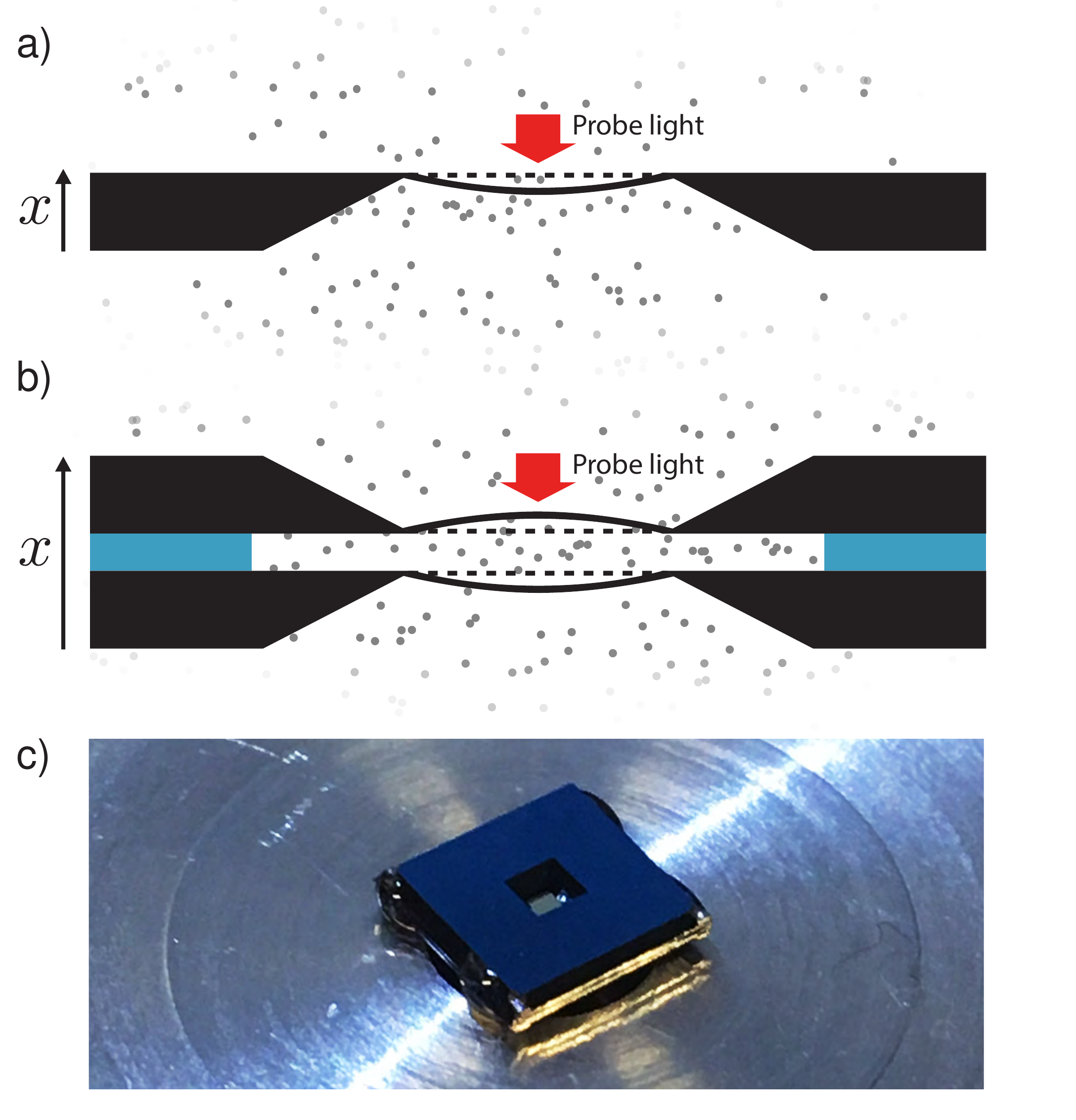}
\caption{Schematic sideviews of single-membrane (a) and double-membrane (b) resonators. c) Photograph of a double-membrane resonator.}
\label{fig1}
\end{figure}

A schematic of the suspended membrane drums used in this work is shown in Fig.~\ref{fig1}. A pair of high tensile stress ($\sim715$~MPa), square silicon nitride membranes (thickness $t=92$~nm, lateral dimension $a=500$~$\mu$m) on a 5 mm-square silicon frame (Norcada Inc., Canada) and separated by a spacer with thickness $d=8.5$~$\mu$m, constitute the air-filled cavity. The membranes exhibit fundamental square drummode frequencies $\omega/(2\pi)\sim 720$~kHz with vacuum quality factors of a few $10^5$~\cite{Nair2017}. The vibrational noise spectrum of the resonators are determined by injecting the light from an external cavity diode laser (wavelength 900-920 nm) into a 6 mm-long linear Fabry-Perot interferometer consisting of the double-membrane resonator and a 50\% transmittivity beamsplitter and analyzing the transmission signal measured by a fast photdetector with a spectrum analyzer. We investigate squeeze film effects on these structures in the range $10^{-5}$-10 mbars. 

Different regimes for the dynamics of the resonator in the fluid can be distinguished based on the Knudsen number, which is the ratio of the gas molecule mean-free path and the typical structure dimension. High Knudsen numbers ($\gtrsim 10$) correspond to the free molecular -- or rarefied air -- regime, in which kinetic damping caused by molecules bouncing off the moving film leads to a friction force proportional with pressure~\cite{Christian1966}, while at low Knudsen numbers ($\lesssim 0.01$) the gas acts as a viscous fluid and damping becomes independent of pressure, with a transition regime in between. These regimes have been investigated with a number of suspended resonators spanning a broad range of sizes and frequencies~\cite{Prak1991,Blom1992,Legtenberg1994,Dohn2005,Vignola2006,Li2007,Verbridge2008,Khine2009, Svitelskiy2009,Ekinci2010}. 
Since the Knudsen number is here of order unity at $\sim$~mbar pressures, the dynamics happen in the transition regime between the free molecular and the viscous regimes. In the latter, the strengths of the elastic and damping forces depend on the dimensionless squeeze number~\cite{Blech1983}, $\sigma=12\mu a^2\omega/pd^2$, where $\mu$ is the fluid viscosity and $p$ the pressure. At atmospheric pressure, $\sigma\sim 40$ for our resonators, which suggests an essentially elastic squeeze film force and negligible squeeze film damping. In the free molecular regime, as discussed in~\cite{Suijlen2009,Ekinci2010}, a more relevant quantity to discuss the nature of the trapping of the fluid is the product $\omega\tau$ of the mechanical frequency by the molecular diffusion time to equalize the pressure inside and outside the gap. Estimating $\tau$ for our structures as in~\cite{Suijlen2009} gives $\omega\tau\sim 80$, also predicting an essentially elastic squeezed film force in this regime.

\begin{figure}
\includegraphics[width=\columnwidth]{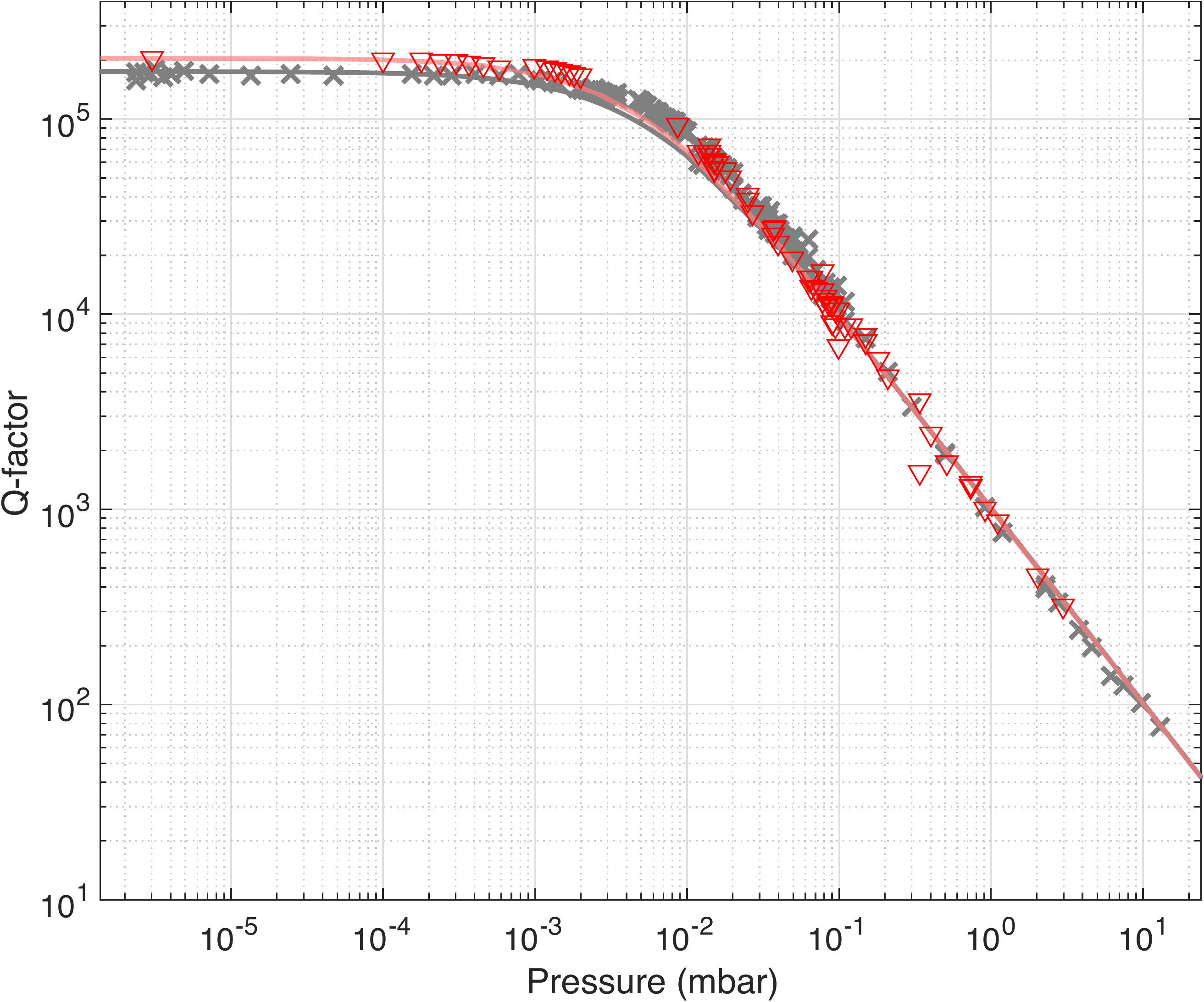}
\caption{Variation of the fundamental mode mechanical quality factor with pressure for a single resonator (crosses) and a double-membrane resonator (triangles). The lines show the theoretical predictions including intrinsic and kinetic damping.}
\label{fig2}
\end{figure}

Figure~\ref{fig2} shows the variations of the mechanical quality factor of the fundamental mode when the pressure $P$ in the vacuum chamber is varied for a single membrane (Fig.~\ref{fig1}a) and for a double-membrane resonator (Fig.~\ref{fig1}b). The Q-factors were measured by ringdown spectroscopy in the range $10^{-6}$-$10^{-1}$~mbar or by Lorentzian fits to the thermal noise spectrum in the range $10^{-2}$-10~mbars. These variations are well-reproduced by the expression $Q=1/(Q_{\textrm{i}}^{-1}+Q_{\textrm{air}}^{-1})$, where $Q_{\textrm{i}}$ is the intrinsic (vacuum) Q-factor and $Q_{\textrm{air}}=\omega/\gamma_{\textrm{air}}$  is the Q-factor value predicted when kinetic damping dominates in the free molecular regime~\cite{Christian1966} with a damping rate \begin{equation}
\gamma_{\textrm{air}}=4\sqrt{\frac{2}{\pi}}\sqrt{\frac{M}{RT}}\frac{P}{\rho t}
\label{eq:gamma}
\end{equation}
with $\rho=2700$ kg/m$^3$ being the density of silicon nitride, $M$ the molar mass of air, $R$ the ideal gas constant and $T$ the temperature. The observation that single- and double-membrane resonators exhbit the same (kinetic) damping show that squeeze film damping is negligible, as expected.

\begin{figure}
\includegraphics[width=\columnwidth]{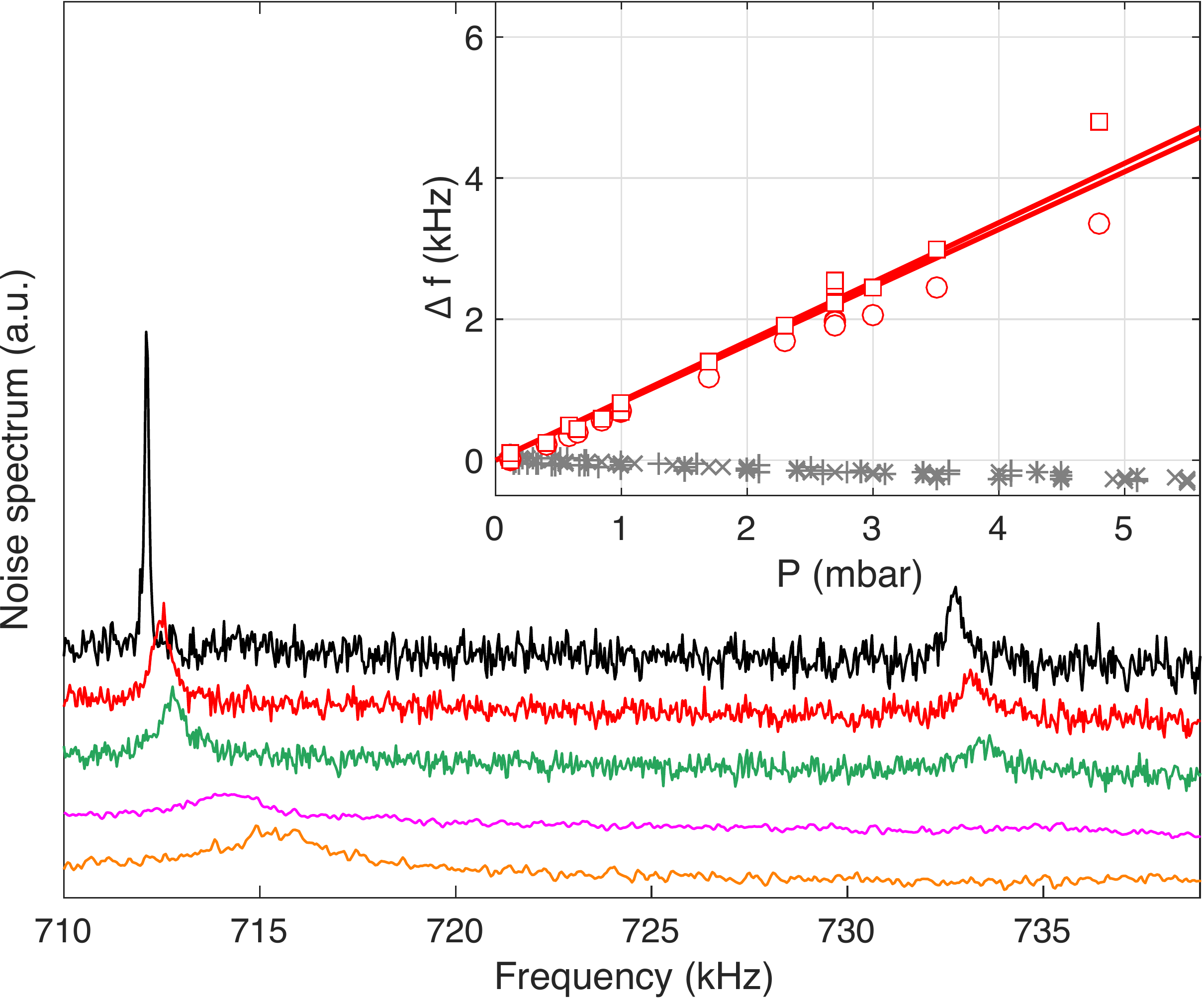}
\caption{Noise spectra for a double-membrane resonator with well-separated frequency modes, at different pressures ranging from 0.18 to 5.1 mbar from top to bottom (the spectra are offset vertically for visibility). The inset shows the resonance frequency shifts as a function of pressure for single membranes (crosses) and the membranes of an array (circles and squares).  The lines show the results of the squeeze film predictions.}
\label{fig3}
\end{figure}

Figure~\ref{fig3} shows typical vibrational mode spectra for a double-membrane resonator at different pressures, as well as the variations of the resonant frequency of the fundamental mode with pressure for both single- and double-membrane resonators. The resonance frequencies were determined via a Lorentzian fit to the spectrum and the shift measured with respect to the low-pressure value. A negative frequency shift, $\Delta f=-52$ Hz/mbar, is observed for single membranes, in accordance with the fact that the effective mass of the displaced fluid has to be added to the resonator mass (the shift due to the decrease in Q is still negligible at these pressures). In contrast, the double-membrane resonator modes at 712 and 733 kHz exhibit much larger and positive linear frequency shifts. Indeed, as the pressure is increased, the isothermal spring constant of the gas $k_{\textrm{air}}=Pa^2/d$ is added to the natural spring constant, $k=m\omega^2$, where $m$ is the oscillator mass. When $k_{\textrm{air}}\ll k$ this results in a positive linear frequency shift for the two modes of 842 and 818 Hz/mbar, as is corroborated by the experimental observations.

\begin{figure}
\includegraphics[width=\columnwidth]{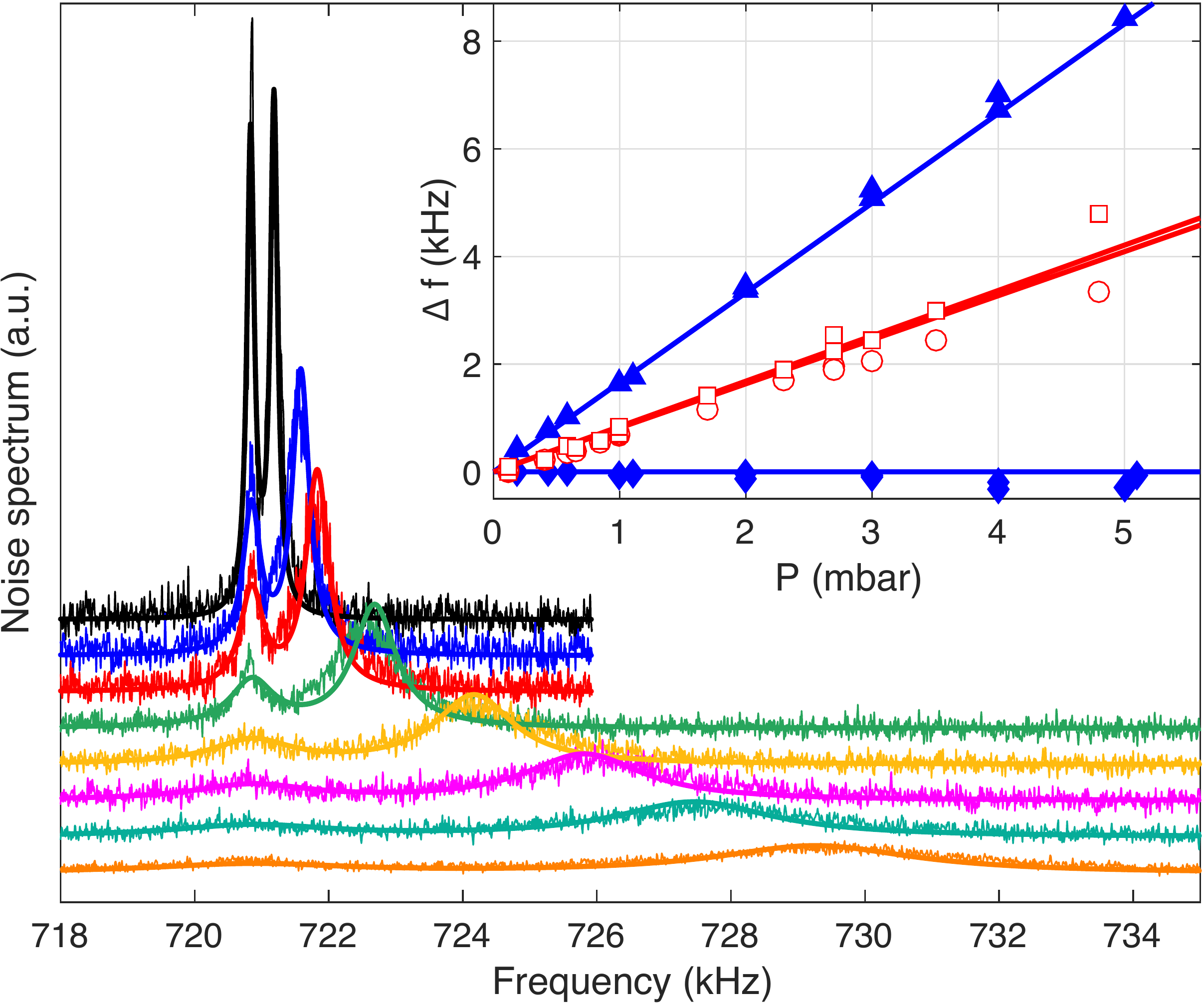}
\caption{Noise spectra for a double-membrane resonator with near degenerate frequency modes. The solid lines show the result of a global fit to the theoretical model. The inset show the hybridized mode frequency shifts as a function of pressure (full symbols), while the frequency shifts for the resonator with well-separated frequency modes used in Fig.~\ref{fig3} (open symbols) are shown again for comparison.}
\label{fig4}
\end{figure}

The results presented so far concerned double-membrane resonators with well-separated mode frequencies. Interestingly, due to the high degree of uniformity exhibited by membranes from the same fabrication batch, several of the assembled arrays displayed near-degenerate frequency modes. In that case, not only does the gas modify mechanical damping and resonant frequencies, but it can also couple the vibrational modes of each membrane when the air-induced shift becomes of the order of the mode frequency separation. This behavior can be observed in Fig.~\ref{fig4}, which shows typical noise spectra of a double-membrane resonator for which the fundamental modes of each membrane are separated by only 0.2~kHz in vacuum. As the pressure is increased, the modes are clearly seen to hybridize into a ``dark" mode whose spring constant remains essentially unchanged and a ``bright" mode whose spring constant is effectively doubled as compared to the non-interacting mode scenario. The spectra can be well-reproduced by a simple coupled oscillator model, in which two modes with natural frequencies $\omega_i$ and vacuum Q-factors $Q_i=\omega_i/\gamma_i$ ($i=1,2$) experience a kinetic damping friction force and a squeeze film-mediated elastic coupling, both proportional to the pressure. The fluctuations of the membrane displacements around equilibrium in Fourier space, $x_1(\omega)$ and $x_2(\omega)$, are then given by
\begin{eqnarray}
\left[\omega_1^2-\omega^2-i\omega(\gamma_1+\gamma_{\textrm{air}})\right]x_1 +k_{\textrm{air}}(x_1-x_2) &=F^{\textrm{th}}_1+F^{\textrm{air}}_1,\\
\left[\omega_2^2-\omega^2-i\omega(\gamma_2+\gamma_{\textrm{air}})\right]x_2 +k_{\textrm{air}}(x_2-x_1)&=F^{\textrm{th}}_2+F^{\textrm{air}}_2,
\end{eqnarray}
where $F_{1,2}^{\textrm{th}}$ and $F_{1,2}^{\textrm{air}}$ are thermal and kinetic damping noise Langevin forces. To measure the noise spectra, the length of the interferometer is stabilized at the maximum slope of its transmission and the wavelength of the incoming light is varied so at to maximize the optomechanical response, and thereby the noise spectra amplitudes, of the compound three-mirror system~\cite{Genes2017}. The measured noise spectrum thus corresponds to that of an \textit{a priori} unknown linear combination of the membrane displacements $a[\cos(\theta)x_1+\sin(\theta)x_2]$, which depends on the modematching of the interferometer, the overlap between the optical beam and the mechanical modes, the distances between the mirrors and the operating wavelength. The results of a global fit with the model, with $a$ and $\theta$ as free parameters and with $\gamma_{\textrm{air}}$ and $k_{\textrm{air}}$ as given by the theoretical predictions, are shown in Fig.~\ref{fig4} and are seen to well reproduce the measured spectra. The independently measured resonance frequency shifts (inset of Fig.~\ref{fig4}) from Lorentzian fits to the spectra, are likewise in good agreement with the theoretically expected shifts of $\simeq 0$ and 1664 Hz/mbar.

To conclude, squeeze film effects in micromechanical membrane cavities have been investigated in the molecular and quasi-molecular regimes and squeeze film-induced hybridization between modes of distinct resonators has been observed. While the intermembrane spacing was fixed in this work to 8.5 $\mu$m, it could easily be reduced in order to increase the magnitude of the squeeze film effects. Appropriately choosing the membrane dimensions and frequencies would allow for optimizing such simple gas coupled micromechanical cavities for either pressure sensing or for frequency shift gas sensing (for monitoring gas density change or adsorption mass loading) applications ~\cite{Suijlen2009}.

We acknowledge support from the Velux Foundations, Carlsbergfondet and the Danish Council for Independent Research.

\bibliography{airdampingarraybib.bib}

\end{document}